\documentclass[letterpaper,10pt]{article} 

\usepackage{opticameet3} %

\usepackage[acronym]{glossaries}
\usepackage{bm}
\usepackage{siunitx}
\usepackage{upgreek}
\usepackage{nicefrac}
\usepackage{kitcolors}
\usepackage{pgfplots}
\usepackage{units}
\usepackage{subcaption}
\usepackage{mathtools}
\usepackage{mleftright}
\usepackage[font={footnotesize}]{caption}

\usepackage{todonotes}

\usetikzlibrary{fit, positioning, calc, angles, arrows.meta}
\pgfplotsset{compat=1.17}

\newacronym{bicm}{BICM}{bit interleaved coded modulation}
\newacronym{bps}{BPS}{blind phase search}
\newacronym{rpn}{RPN}{residual phase noise}
\newacronym{awgn}{AWGN}{additive white Gaussian noise}
\newacronym{gs}{GS}{geometrical shaping}
\newacronym{qam}{QAM}{quadrature amplitude modulation}
\newacronym{snr}{SNR}{signal-to-noise ratio}
\newacronym{bce}{BCE}{binary cross entropy}
\newacronym{bmi}{BMI}{bitwise mutual information}
\newacronym{gcs}{GCS}{geometric constellation shaping}
\newacronym{gmi}{GMI}{generalized mutual information}
\newacronym{mi}{MI}{mutual information}
\newacronym{e2e}{E2E}{end-to-end}
\newacronym{cpe}{CPE}{carrier phase estimation}
\newacronym{cps}{CPS}{carrier phase synchronization}
\newacronym{llr}{LLR}{log likelikood ratio}
\newacronym{nn}{NN}{neural network}
\newacronym{tx}{Tx}{transmitter}
\newacronym{rx}{Rx}{receiver}
\newacronym{fec}{FEC}{forward error correction}
\newacronym{bmd}{BMD}{bit-metric decoder}
\newacronym{ff-nn}{FF-NN}{feed-forward neural network}
\newacronym{ber}{BER}{bit error rate}
\newacronym{dsp}{DSP}{digital signal processing}

\newcommand\blfootnote[1]{%
  \begingroup
  \renewcommand\thefootnote{}\footnote{#1}%
  \addtocounter{footnote}{-1}%
  \endgroup
}

\colorlet{KITColor1}{kit-blue100}
\colorlet{KITColor2}{kit-orange100}
\colorlet{KITColor3}{kit-red100}
\colorlet{KITColor4}{kit-purple100}
\colorlet{KITColor5}{kit-cyanblue100}

\let\Re\relax
\let\Im\relax
\DeclareMathOperator{\Re}{Re}
\DeclareMathOperator{\Im}{Im}
\DefineSpuriousJournalWord{on}

\makeatletter
\DeclareRobustCommand{\rvdots}{%
  \vbox{
    \baselineskip4\p@\lineskiplimit\z@
    \kern-\p@
    \hbox{.}\hbox{.}\hbox{.}
  }}
\makeatother

\newcommand\authormark[1]{\textsuperscript{#1}}

\usepackage[colorlinks=true,bookmarks=false,citecolor=blue,urlcolor=blue]{hyperref} %

\begin{document}

\title{Geometric Constellation Shaping with Low-complexity Demappers for Wiener Phase-noise Channels}

\copyrightyear{2023}

\author{Andrej Rode\authormark{*} and Laurent Schmalen}

\address{Communications Engineering Lab (CEL), Karlsruhe Institute of Technology, 76187 Karlsruhe, Germany}

\email{\authormark{*}\texttt{rode@kit.edu}} %

\begin{abstract}
We show that separating the in-phase and quadrature component in optimized, machine-learning based demappers of optical communications systems with geometric constellation shaping reduces the required computational complexity whilst retaining their good performance.%
\end{abstract}

\glsresetall
\suppressfloats[t] %
\section{Introduction}\vspace*{-1ex}
In recent years, deep learning and auto-encoders for \gls{e2e} optimization of optical communication systems\cite{karanovEndtoEndDeepLearning2018, uhlemannDeeplearningAutoencoderCoherent2020} have gained much attention. Specifically, \gls{gcs} for the Wiener phase noise channel is a promising approach to improve spectral efficiency\cite{dzieciolGeometricShaping2D2021,jovanovicEndtoendLearningConstellation2022a}. Often, blocks of the \gls{dsp} chain have been left out of the optimization via auto-encoders, either due to an increase in computational complexity or problems which they introduce in the differentiability of the channel\cite{yankovRecentAdvancesConstellation2022}, required for optimization. One such \gls{dsp} block is the \gls{bps}\cite{pfauHardwareefficientCoherentDigital2009}, which is a popular algorithm for blind, feed-forward, \gls{cps} in high-rate coherent optical communication receivers. In \cite{rodeGeometricConstellationShaping2022a}, the authors implemented a differentiable \gls{bps} to enable a fully differentiable channel model for the optimization of \gls{gcs} in the presence of Wiener phase noise.

Previous works \cite{jovanovicEndtoendLearningConstellation2021a, jovanovicEndtoendLearningConstellation2022a, rodeGeometricConstellationShaping2022a, rodeOptimizationGeometricConstellation2022b} successfully applied \gls{gcs} to improve the spectral efficiency and robustness of optical communication systems with \gls{bps}.\blfootnote{This work has received funding from the European Research Council (ERC)
under the European Union’s Horizon 2020 research and innovation programme
(grant agreement No. 101001899).} %
 To harness the full potential of a neural receiver, previous works use \gls{nn}-based demappers with multiple fully connected layers. This allows the system to train the neural network to fully adapt to the channel. Due to the high number of weights and, subsequently, a high number of arithmetic operations per received symbol, such \gls{nn}-based demappers add computational complexity to the receiver. In a related work, it has been explored to reduce the computational complexity by limiting the \gls{nn}-based demapper to a single hidden layer\cite{schaedlerNeuralNetworkBasedSoftDemapping2020}. In this work, we present an approach to reduce the computational complexity by separating the in-phase and quadrature components at the \gls{nn}-based demapper and reducing the size of the demapper \glspl{nn} in both width, and depth. We compare our approach to a simple reduction of the \gls{nn}-based demapper in width and size. The separation of the in-phase and quadrature components follows the implementation of demappers for classical square QAM.\vspace*{-2ex}

\section{System Model}\vspace*{-4ex}
\begin{figure}[!h]
    \centering
    \newsavebox\neuralnetwork
\sbox{\neuralnetwork}{%
		\begin{tikzpicture}[
      >=stealth,
      scale=.9,
      every node/.append style={transform shape},
      remember picture,
      ]%
      \tikzset{Source1b/.style={rectangle, draw=black, thick, minimum width=0.05cm, minimum height=1.2cm, rounded corners=0.5mm}}
      \tikzset{Source3/.style={rectangle, draw, thick, minimum width=0.6cm, minimum height=0.45cm, rounded corners=0.5mm}}

      \tikzset{OnehotNode/.style={circle, thick, draw,minimum width=0.1cm}}
      \tikzset{ReLUNode/.style={circle,thick,draw,fill=black!10!white}}
      \tikzset{MZMNode/.style={circle,thick,draw,fill=black!30!white}}
   \node (serial) at (3.6,0) {};
	  \def\offseta{4.5}
	  \def\offsetb{0.5cm}
	  \def\offsetc{15cm}
    \def\k{0}
    \def\ki{1}
        \node [OnehotNode] (S\ki1) at ($(0,0.5)+(0,\k)$) {};
        \node [OnehotNode] (S\ki2) at ($(0,-0.5)+(0,\k)$) {};
		\node at ($(0,0)+(0,\k)$) {$\rvdots$};
		\draw [->] ($(S\ki1)+(-0.4cm,0)$) -- (S\ki1);
		\draw [->] ($(S\ki2)+(-0.4cm,0)$) -- (S\ki2);

		\node [ReLUNode] (H\ki11) at ($(1,1.5)+(0,\k)$) {};
		\node [ReLUNode] (H\ki12) at ($(1,0.5)+(0,\k)$) {};
		\node [ReLUNode] (H\ki13) at ($(1,-0.5)+(0,\k)$) {};
		\node [ReLUNode] (H\ki14) at ($(1,-1.5)+(0,\k)$) {};
		\node at ($(1,0)+(0,\k)$) [anchor=center] {$\rvdots$};

		\foreach\j in {1,2,3,4}{
		  \draw [->] (S\ki1) -- (H\ki1\j); \draw [->] (S\ki2) -- (H\ki1\j);
		}

		\node [ReLUNode] (H\ki21) at ($(2,1.5)+(0,\k)$) {};
		\node [ReLUNode] (H\ki22) at ($(2,0.5)+(0,\k)$) {};
		\node [ReLUNode] (H\ki23) at ($(2,-0.5)+(0,\k)$) {};
		\node [ReLUNode] (H\ki24) at ($(2,-1.5)+(0,\k)$) {};
		\node at ($(2,0)+(0,\k)$) [anchor=center] {$\rvdots$};

		\foreach\j in {1,2,3,4}{
		  \foreach\i in {1,2,3,4}{
		    \draw [->] (H\ki1\j) -- (H\ki2\i);
		  }
		}

		\node [MZMNode] (M\ki1) at ($(3,1)+(0,\k)$) {};
		\node [MZMNode] (M\ki2) at ($(3,0)+(0,\k)$) {};
		\node [MZMNode] (M\ki3) at ($(3,-1)+(0,\k)$) {};
		\draw [<-] ($(serial.west)+(0,\k)+(0,1)$) -- (M\ki1.east);
		\draw [<-] ($(serial.west)+(0,\k)$) -- (M\ki2.east);
		\draw [<-] ($(serial.west)+(0,\k)+(0,-1)$) -- (M\ki3.east);

		\node at ($(3,-0.5)+(0,\k)$) [anchor=center] {$\rvdots$};
		\node at ($(3,+0.5)+(0,\k)$) [anchor=center] {$\rvdots$};
		\foreach\j in {1,2,3,4}{
		  \foreach\i in {1,2,3}{
		    \draw [->] (H\ki2\j) -- (M\ki\i);
		  }
		}
\end{tikzpicture}%
}

\newsavebox\thinneural
\sbox{\thinneural}{%
		\begin{tikzpicture}[
      >=stealth,
      scale=.9,
      every node/.append style={transform shape},
      remember picture,
      ]%
      \tikzset{Source1b/.style={rectangle, draw=black, thick, minimum width=0.05cm, minimum height=1.2cm, rounded corners=0.5mm}}
      \tikzset{Source3/.style={rectangle, draw, thick, minimum width=0.6cm, minimum height=0.45cm, rounded corners=0.5mm}}

      \tikzset{OnehotNode/.style={circle, thick, draw,minimum width=0.1cm}}
      \tikzset{ReLUNode/.style={circle,thick,draw,fill=black!10!white}}
      \tikzset{MZMNode/.style={circle,thick,draw,fill=black!30!white}}
   \node (serial) at (2.6,0) {};
	  \def\offseta{4.5}
	  \def\offsetb{0.5cm}
	  \def\offsetc{15cm}
    \def\k{0}
    \def\ki{1}
        \node [OnehotNode] (S\ki1) at ($(0,0)+(0,\k)$) {};
		\draw [->] ($(S\ki1)+(-0.4cm,0)$) -- (S\ki1);

		\node [ReLUNode] (H\ki11) at ($(1,1.5)+(0,\k)$) {};
		\node [ReLUNode] (H\ki12) at ($(1,0.5)+(0,\k)$) {};
		\node [ReLUNode] (H\ki13) at ($(1,-0.5)+(0,\k)$) {};
		\node [ReLUNode] (H\ki14) at ($(1,-1.5)+(0,\k)$) {};
		\node at ($(1,0)+(0,\k)$) [anchor=center] {$\rvdots$};

		\foreach\j in {1,2,3,4}{
		  \draw [->] (S\ki1) -- (H\ki1\j);
		}

		\node [MZMNode] (M\ki1) at ($(2,1)+(0,\k)$) {};
		\node [MZMNode] (M\ki2) at ($(2,0)+(0,\k)$) {};
		\node [MZMNode] (M\ki3) at ($(2,-1)+(0,\k)$) {};
		\draw [<-] ($(serial.west)+(0,\k)+(0,1)$) -- (M\ki1.east);
		\draw [<-] ($(serial.west)+(0,\k)$) -- (M\ki2.east);
		\draw [<-] ($(serial.west)+(0,\k)+(0,-1)$) -- (M\ki3.east);

		\node at ($(2,-0.5)+(0,\k)$) [anchor=center] {$\rvdots$};
		\node at ($(2,+0.5)+(0,\k)$) [anchor=center] {$\rvdots$};
		\foreach\j in {1,2,3,4}{
		  \foreach\i in {1,2,3}{
		    \draw [->] (H\ki1\j) -- (M\ki\i);
		  }
		}
\end{tikzpicture}%
}

\tikzset{MUL/.style={draw,circle,append after command={
      [shorten >=\pgflinewidth, shorten <=\pgflinewidth,]
      (\tikzlastnode.north west) edge (\tikzlastnode.south east)
      (\tikzlastnode.north east) edge (\tikzlastnode.south west)
    }
  }
}
\tikzset{ADD/.style={draw,circle,append after command={
      [shorten >=\pgflinewidth, shorten <=\pgflinewidth,]
      (\tikzlastnode.north) edge (\tikzlastnode.south)
      (\tikzlastnode.east) edge (\tikzlastnode.west)
    }
  }
}
\tikzstyle{box}=[rectangle,draw=black, minimum size=7mm, inner sep=2mm, font=\footnotesize, align=center]
\tikzstyle{node}=[circle, draw=black, minimum size=5mm]
\tikzstyle{connection}=[->,>=latex]
\tikzset{%
  cblock/.style    = {draw, thick, rectangle, minimum height = 3em, minimum width = 6em,rounded corners=1mm},
  lblock/.style = {draw,thick,rectangle,minimum height=10em, minimum width=2em,
    rounded corners=1mm},
  operation/.style = {draw, thick, minimum height= 1.5em, circle},
}
\def\aend{0.25}
	\resizebox{0.8\textwidth}{!}{
		\begin{tikzpicture}[font=\huge, thick]
		    \node [lblock, minimum width=12em] (embed) at (0,1) {\LARGE \begin{tabular}{c}
		         Binary  \\
		         $\downarrow$ \\
		         One-Hot
		    \end{tabular}};

        \node [operation, below right=0.5 of embed, append after command={node[fill,circle,minimum size=3pt, inner sep=0pt] at (\tikzlastnode.center){}}] (tx_mult) {};
        \node [lblock, below=1 of embed, minimum width=12em] (txnet) {\LARGE\begin{tabular}{c} Tx-NN\\\usebox{\neuralnetwork}\end{tabular}};
        \node [operation, ADD, right=2.5 of tx_mult] (awgn_add) {};
        \node [below=1 of awgn_add, text depth=1.5em, text height=1.5em] (awgn_param) {$n_k$};

        \node [operation, MUL, right=0.8 of awgn_add] (phase_mult) {};
        \node [below=1 of phase_mult, text depth=1.5em, text height=1.5em] (pn_param) {$\mathrm{e}^{\mathrm{j}\varphi}$};
		\node [above=1.2 of phase_mult] (pn_param_def) {$\Delta\varphi \sim \mathcal{N}(0,\sigma_\upphi^2$)};

        \node [coordinate,right=0.5 of phase_mult] (switch_start) {};
        \node [coordinate,right=0.5 of switch_start] (switch_cent) {};
        \node [coordinate,above=1 of switch_cent] (switch_top) {};
        \node [coordinate,below=1 of switch_cent] (switch_bottom) {};
        \pic [draw, angle radius=15, dotted, <-, >={Latex[scale=1.1]}, very thick] {angle=switch_bottom--switch_start--switch_top};
        \node [cblock, KITColor2, right=0.75 of switch_top, minimum width=10em] (diff_bps) {\LARGE diff. BPS};
        \node [cblock, right=0.75 of switch_bottom, minimum width=10em] (bps) {\LARGE BPS};
        \node [coordinate,right = 1 of bps] (end_switch_bottom) {};
        \node [coordinate,right = 0.75 of diff_bps] (end_switch_top) {};
        \node [coordinate,below = 0.75 of end_switch_top] (end_switch_cent) {};
        \node [coordinate,right = 0.5 of end_switch_cent] (end_switch_end) {};
        \node [coordinate,right = 1 of end_switch_end] (complexrx_split){};
        \node [coordinate,right = 3 of complexrx_split] (complexrx_cent) {};
        \pic [draw, angle radius=15, dotted, ->, >={Latex[scale=1.1]}, very thick] {angle=end_switch_top--end_switch_end--end_switch_bottom};
		\node [lblock, above= 0.25 of complexrx_cent] (complexrx_real) {\LARGE\begin{tabular}{c}{$\Re$-NN}\\\usebox{\thinneural}\end{tabular}};
        \node [lblock, below= 0.25 of complexrx_cent] (complexrx_imag) {\LARGE\begin{tabular}{c}{$\Im$-NN}\\\usebox{\thinneural}\end{tabular}};
		\node [right=0.6 of complexrx_real,align=left] (sink_real) {$\begin{pmatrix*}[l]\hat{L}_{b,1}\\ \hat{L}_{b,2}\\ \vdots\\ \hat{L}_{b,m/2}\end{pmatrix*}$};
  		\node [right=0.6 of complexrx_imag,align=left] (sink_imag) {$\begin{pmatrix*}[l]\hat{L}_{b,m/2 + 1}\\ \hat{L}_{b,m/2 +2}\\ \vdots\\ \hat{L}_{b,m}\end{pmatrix*}$};

		\node [left=1 of embed,align=left] (source) {$\begin{pmatrix*}[l]b_1\\ b_2\\ \vdots\\ b_m\end{pmatrix*}$};
		\draw [-{Latex[scale=1.25]},thick] (source) -- (embed);
		\draw [-{Latex[scale=1.25]},thick] (awgn_param) -- (awgn_add);
		\draw [-{Latex[scale=1.25]},thick] (pn_param) -- (phase_mult);
		\draw [-{Latex[scale=1.25]},thick] (txnet.east) -| (tx_mult);
        \draw [-{Latex[scale=1.25]},thick] (embed.east) -| (tx_mult);
        \draw [-{Latex[scale=1.25]},thick] (tx_mult) -- node[above left]{$x_k$} (awgn_add);
        \draw [-{Latex[scale=1.25]},thick] (awgn_add) -- (phase_mult);
        \draw [-{Latex[scale=1.25]},thick] (phase_mult) -- node[above]{$z_k$} (switch_start) -- (switch_top);
        \draw [{Circle[open]}-, thick] (switch_top) -- (diff_bps);
        \draw [{Circle[open]}-,thick] (switch_bottom) -- (bps);
        \draw [-{Circle[open]},thick] (bps) -- (end_switch_bottom);
        \draw [-{Circle[open]},thick] (diff_bps) -- (end_switch_top);
        \draw [-{Latex[scale=1.25]},thick] (end_switch_top) -- (end_switch_end) -- node[above,near start,name=hat_xk]{$\hat{x}_k$}(complexrx_split); %
        \draw [-{Latex[scale=1.25]},thick] (complexrx_split) |- (complexrx_real);
        \draw [-{Latex[scale=1.25]},thick] (complexrx_split) |- (complexrx_imag);
        \draw [-{Latex[scale=1.25]},thick] (complexrx_real) -- (sink_real);
        \draw [-{Latex[scale=1.25]},thick] (complexrx_imag) -- (sink_imag);

		    \node (rectch) [fit={($(awgn_add.north west)+(-1,3)$) ($(end_switch_end.south east)+(1.2,-3.2)$)}, draw, rounded corners=6pt, dashed, KITColor1,inner sep=0pt,rectangle] {};
		    \node at (rectch.north) [anchor=north,KITColor1,above] {Auto-encoder channel};
		\end{tikzpicture}}
    \captionof{figure}{Block diagram of the bitwise auto-encoder system with a separated demapper \gls{nn}}\vspace*{-3ex}%
    \label{fig:system_model}
\end{figure}
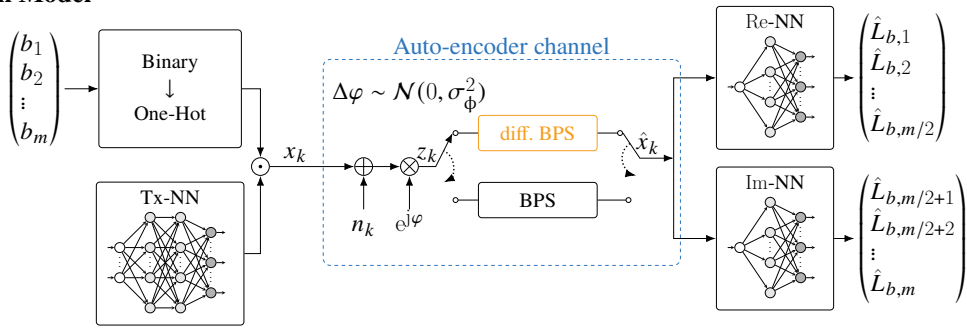
In our proposed approach, we model an \gls{e2e} optical transmission system at the symbol rate. The transmitter is implemented as an array of trainable weights for the in-phase and quadrature component, which is the Tx-NN in the block diagram of the system in Fig.~\ref{fig:system_model}. Following our previous works\cite{rodeGeometricConstellationShaping2022a,rodeOptimizationGeometricConstellation2022b}, we implement \gls{gcs} following the bitwise auto-encoder approach with \gls{bicm}\cite{cammererTrainableCommunicationSystems2020}. In the bitwise auto-encoder, bit vectors of length $m$ are mapped to a complex-valued constellation of size $M = 2^m$. This is accomplished by taking the dot-product between the one-hot vector and the vector of complex constellation points at the output of the Tx-NN.
The complex transmit symbols are impaired by \gls{awgn} with standard deviation $\sigma_{n}$ and Wiener phase noise with the standard deviation $\sigma_{\upphi}$ of the process increments. Both values are selected to represent a specific \gls{snr} and laser linewidth at the given symbol rate. The impaired and distorted complex received symbols are then sent through a \gls{cps}, which is implemented by means of \gls{bps}. During training, the operation of \gls{bps} is approximated by the differentiable \gls{bps}\cite{rodeGeometricConstellationShaping2022a} with a temperature parameter, which is gradually reduced from \num{1.0} to \num{0.001} to approximate the non-differentiable \gls{bps} more closely at the end of the training.
After the \gls{bps}, the received and corrected symbols $\hat{x}_k$ are separated in in-phase and quadrature and passed through individual \glspl{nn}, $\Re$-NN and $\Im$-NN, each with one hidden layer, as depicted in Fig.~\ref{fig:system_model}. Each \gls{nn} then returns $m/2$ \glspl{llr}, which are used to calculate the loss function, the \gls{bce}. The \gls{bce} is used as a loss function to maximize the \gls{bmi}\cite{cammererTrainableCommunicationSystems2020}. To validate the \gls{e2e} performance in the context of an implementable system, the non-differentiable \gls{bps} algorithm is used in the validation runs. Parameters of the \gls{bps} are selected as $N_{\text{angles}} = 60$ test angles and a window size of $N = 120$ for averaging.\vspace*{-1ex}

\section{Complexity Assessment of Neural Network-based Demappers}\vspace*{-1ex}
To compare the complexity of our proposed approach, we compare the computational complexity of classical calculation of \glspl{llr} with its computation via neural networks. As pointed out in \cite{schaedlerNeuralNetworkBasedSoftDemapping2020}, soft demapping to obtain \glspl{llr} can be performed with a log-MAP demapper with \gls{awgn} assumption or with an \gls{nn}, formulated respectively with
\begin{equation}
    L_{i}\mleft(y\mright) = 
    \ln\left(
    \frac
    {\displaystyle \sum_{x \in \mathcal{X}_i^0}{\exp\left({\frac{| y - x |^2 }{N_0}}\right)}}
    {\displaystyle \sum_{x \in \mathcal{X}_i^1} \exp\left({\frac{|y - x |^2 }{N_0}}\right)}
    \right)
\qquad
\text{and}
\qquad
    \bm{L}\mleft(y\mright) = \left(\tau\left(\bm{y}^T\bm{W}_1 + \bm{b}_1\right)\right)^T\bm{W}_2 + \bm{b}_2.
    \label{eq:llr_calculation}
\end{equation}
$\mathcal{X}_i^b$ is the set of symbols with a binary label that takes on value $b$ at the $i$th bit position. The computational complexity for calculating an \gls{llr} vector for an $M$-\gls{qam} constellation consists of calculating $M$ complex subtractions and $M$ complex multiplications. Afterwards, $2 m = 2\log_2(M)$ sums with $M/2$ elements each have to be accumulated and the exponential has to be calculated with another $m$ divisions and calculation of $\ln$ following these operations. %
The complexity of \glspl{nn} can be assessed in terms of the vector-matrix multiplication needed for each layer. Each layer in an \gls{nn} needs $nm $ multiplications, $nm$ additions, and $m$ executions of the non-linear activation function $\tau$, where $n$ is the input width and $m$ is the output width. For an \gls{nn}-based demapper with input width $k$, $\ell$ hidden layers and $m$ outputs, the number of multiplications $L_{\text{NN,mult}}$ required is
    $L_{\text{NN,mult}} = kn + (\ell-1)n + nm$.
 We compare the computational complexity between the Gaussian demapper and the \gls{nn}-based demapper via the number of real-valued multiplications required to calculate the \glspl{llr} per received symbol. The Gaussian demapper in \eqref{eq:llr_calculation} needs $4M$ real-valued multiplications per received symbol. If we relate this to the \gls{nn}-based demapper in \eqref{eq:llr_calculation}, which operates on a complex input and a single hidden layer of size $n$, and solve for $n$ we obtain
$n =\frac{4\left(2^m\right)}{m +2}$.%
We assume that one complex multiplication can be expressed with 4 real-valued multiplications.
For $m=6$, a neural network with a single hidden layer of width $n=32$ needs a similar number of real-valued multiplications compared to a soft demapper with Gaussian noise assumption. For two \glspl{nn} processing the received symbol separately in in-phase and quadrature components, the number of real-valued multiplications is the same for the input and the output layer compared to a single \gls{nn} with complex input.\vspace*{-1ex} 

\section{Results}\vspace*{-1ex}
When constraining the demappers by separation to in-phase and quadrature components we obtain \gls{gcs}-formats resembling classical square \gls{qam} constellations. One half of the bit vector is mapped to amplitudes in the in-phase and the other half to the amplitudes in quadrature.
Even though the constellation resembles square \gls{qam}, we obtain constellations that are optimized to aid the \gls{bps} algorithm in recovering and correcting the carrier phase without the aid of any pilot sequence. This can be inferred from the non-uniform distances between the amplitudes and asymmetrical placement of corner symbols in the constellation diagram in Fig~\ref{fig:constellations}-a). At the same time, the demappers for in-phase and quadrature components are constrained to only one hidden layer and \num{8} weights. Compared to other constellation diagrams trained with low-complexity demappers which do not treat in-phase and quadrature separately, as depicted in Fig.~\ref{fig:constellations}-b) and c), we observe a better separation of constellation points in Fig.~\ref{fig:constellations}-a). In Fig.~\ref{fig:performance} .we compare the performance of the constellations from Fig.~\ref{fig:constellations} and their corresponding neural demappers in terms of \gls{bmi}. All systems have been trained on the same \gls{snr} of \SI{17}{dB} and a laser linewidth of $\Delta\nu = \SI{100}{kHz}$. In Fig.~\ref{fig:performance}-a), the performance is depicted for a fixed linewidth of $\Delta\nu = \SI{100}{kHz}$ and varying \gls{snr}. In Fig.~\ref{fig:performance}-b), the performance is shown if the \gls{snr} is constant at \SI{17}{dB} and the laser linewidth is changed. In both plots, the constellation with the separated demapper is able show a similar performance compared to the constellation and neural demapper presented in \cite{rodeGeometricConstellationShaping2022a}, which uses \num{3} hidden layers with \num{128} nodes per layer. In comparison, the separated demapper uses only \num{1} hidden layer and \num{8} nodes for $\Im$-NN and $\Re$-NN. We explain this matched performance with the fact that the transmit constellation, which we obtained with the separated demapper, forms a real Gray mapping, while the one obtained with the fully-connected demappers only approximately obtain a Gray mapping. Especially when using a lower number of weights in the neural demapper, the performance of the fully-connected demapper drops significantly compared to the separated demapper.\vspace*{-1ex}

\section{Conclusion}\vspace*{-1ex}
By separating the neural demapper in in-phase and quadrature, we are able to apply \gls{gcs} to communication systems impaired by Wiener phase noise and increase spectral efficiency and robustness of these constellations while keeping the implementation complexity of the neural demapper low. This work shows how communication systems that are implemented and optimized with neural networks can be simplified to obtain a lower complexity and how to implement them in a complexity-constrained environment.\vspace*{-2ex}

\begin{figure}[!t]
  \centering
\begin{subfigure}{\textwidth/3}\vspace*{-1.5ex}
  \centering
  \begin{tikzpicture}%
    \begin{axis}[
      xlabel={$\Re\left\{x_k\right\}$},
      ylabel={$\Im\left\{x_k\right\}$},
      height=\textwidth,
      width=\textwidth,
      xmin=-1.8,
      xmax=1.8,
      ymin=-1.8,
      ymax=1.8,
      grid=both,
      ]
      \addplot[
      only marks,
      mark=*,
      mark size=1pt,
      color=KITColor1,
      coordinate style/.from={black,scale=0.5,xshift=5pt},
      nodes near coords,
      point meta=explicit symbolic,
      ]
      table[col sep=tab, meta=label]
      {data/constellation_separated_100kHz.txt};
    \end{axis}
    \node[below = 3mm,xshift=-0.7cm] {a)};
  \end{tikzpicture}
\end{subfigure}%
\begin{subfigure}{\textwidth/3}\vspace*{-1.5ex}
  \centering
  \begin{tikzpicture}%
    \begin{axis}[
      xlabel={$\Re\left\{x_k\right\}$},
      ylabel={$\Im\left\{x_k\right\}$},
      height=\textwidth,
      width=\textwidth,
      xmin=-1.8,
      xmax=1.8,
      ymin=-1.8,
      ymax=1.8,
      grid=both,
      ]
      \addplot[
      only marks,
      mark=*,
      mark size=1pt,
      color=KITColor1,
      coordinate style/.from={black,scale=0.5,xshift=5pt},
      nodes near coords,
      point meta=explicit symbolic,
      ]
      table[col sep=tab, meta=label]
      {data/constellation_w16.txt};
    \end{axis}
    \node[below = 3mm,xshift=-0.7cm] {b)};
  \end{tikzpicture}
\end{subfigure}%
\begin{subfigure}{\textwidth/3}\vspace*{-1.5ex}
  \centering
  \begin{tikzpicture}%
    \begin{axis}[
      xlabel={$\Re\left\{x_k\right\}$},
      ylabel={$\Im\left\{x_k\right\}$},
      height=\textwidth,
      width=\textwidth,
      xmin=-1.8,
      xmax=1.8,
      ymin=-1.8,
      ymax=1.8,
      grid=both,
      ]
      \addplot[
      only marks,
      mark=*,
      mark size=1pt,
      color=KITColor1,
      coordinate style/.from={black,scale=0.5,xshift=5pt},
      nodes near coords,
      point meta=explicit symbolic,
      ]
      table[col sep=tab, meta=label]
      {data/constellation_w8.txt};
    \end{axis}
    \node[below = 3mm,xshift=-0.7cm] {c)};
  \end{tikzpicture}
\end{subfigure}\vspace*{-1ex}%
\caption{Constellation diagrams for $M = 64$, \SI{17}{dB} SNR and laser linewidth \SI{100}{kHz}, a) is trained with a separated low-complexity demapper with $n=8$, b) is trained with a low-complexity demapper with $n = 16$ and c) is trained with a low-complexity demapper with $n=8$}\vspace*{-2ex}
\label{fig:constellations}
\end{figure}
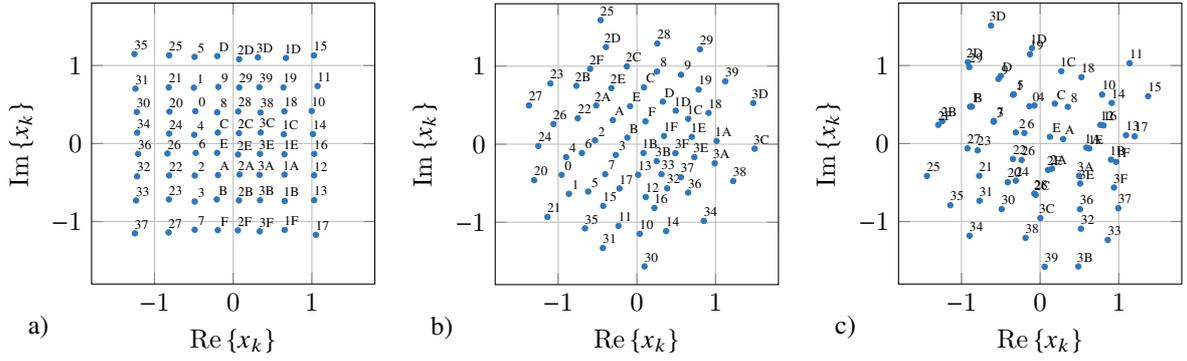
\begin{figure}
\begin{subfigure}{\textwidth/2}
\centering%
\begin{tikzpicture}
\pgfplotsset{
bps/.style ={KITColor2,dashed,mark=*, mark options={solid}},
bps_sep/.style ={KITColor1,solid,mark=square*, mark options={solid}},
bps_full_16/.style={KITColor3,dotted,mark=triangle*, mark options={solid}},
bps_full_8/.style={KITColor4,dotted,mark=diamond*, mark options={solid}},
qam/.style={KITColor5, dashdotted, mark=pentagon*, mark options={solid}}
}
  \begin{axis}[
    xlabel={SNR (\si{dB})},
    ylabel={BMI (\unitfrac{bit}{symbol})},
    ymin=3.5,
    ymax=6,
    xmin=14,
    xmax=25,
    grid=both,
    width=0.9\textwidth,
    height=0.65\textwidth,
    mark size=1pt,
    legend style={at={(axis cs:25
    ,3.5)},anchor=south east,nodes={transform shape},font=\footnotesize},
    label style={font=\small},
    legend cell align={left},
    every axis plot/.append style={very thick, mark size=1.5pt},
    ]
    \addplot[
    bps_sep,error bars/.cd,x dir=none, y dir=both, y explicit, 
    ]
    table[x=snr, y=mean, col sep=space, y error=stddev]
    {data/separated_100kHz.txt};
        \addplot[
    bps,error bars/.cd,x dir=none, y dir=both, y explicit, 
    ]
    table[x=snr, y=mean, col sep=space, y error=stddev]
    {data/ofc22_100kHz.txt};
        \addplot[
    bps_full_16,error bars/.cd,x dir=none, y dir=both, y explicit, 
    ]
    table[x=snr, y=mean, col sep=space, y error=stddev]
    {data/nn_16w_100kHz.txt};
            \addplot[
    bps_full_8,error bars/.cd,x dir=none, y dir=both, y explicit, 
    ]
    table[x=snr, y=mean, col sep=space, y error=stddev]
    {data/nn_8w_100kHz.txt};
    \legend{separated, full-size\cite{rodeGeometricConstellationShaping2022a}, full $n=16$, full $n=8$,square QAM}
    \node [font=\small, anchor=south west] at (axis cs:16,3.6) {$\Delta\nu = \SI{100}{kHz}$};
    \end{axis}
    \node[below = 3mm,xshift=-0.7cm] {a)};
    \end{tikzpicture}
\end{subfigure}%
\begin{subfigure}{\textwidth/2}
\centering
\begin{tikzpicture}
\pgfplotsset{
bps/.style ={KITColor2,dashed,mark=*, mark options={solid}},
bps_sep/.style ={KITColor1,solid,mark=square*, mark options={solid}},
bps_full_16/.style={KITColor3,dotted,mark=triangle*, mark options={solid}},
bps_full_8/.style={KITColor4,densely dotted,mark=diamond*, mark options={solid}},
qam/.style={KITColor5, dashdotted, mark=pentagon*, mark options={solid}}
}
  \begin{axis}[
    xlabel={Laser linewidth (\si{kHz})},
    ylabel={BMI (\unitfrac{bit}{symbol})},
    ymin=4.5,
    ymax=5.4,
    xmin=50,
    xmax=600,
    grid=both,
    width=0.9\textwidth,
    height=0.65\textwidth,
    mark size=1pt,
    legend style={at={(axis cs:600,4.5)},anchor=south east,nodes={transform shape},font=\footnotesize},
    label style={font=\small},
    legend cell align={left},
    every axis plot/.append style={very thick, mark size=1.5pt},
    y tick label style={
        /pgf/number format/.cd,
        fixed,
        fixed zerofill,
        precision=1,
        /tikz/.cd
    }
    ]
    \addplot[
    bps_sep,error bars/.cd,x dir=none, y dir=both, y explicit, 
    ]
    table[x expr={\thisrow{linewidth}/1000}, y=mean, col sep=space, y error=stddev]
    {data/separated_17dB.txt};
    \addplot[
    bps,error bars/.cd,x dir=none, y dir=both, y explicit, 
    ]
    table[x expr={\thisrow{linewidth}/1000}, y=mean, col sep=space, y error=stddev]
    {data/bmi_new_17dB_OFC.txt};
    \addplot[
    bps_full_16,error bars/.cd,x dir=none, y dir=both, y explicit, 
    ]
    table[x expr={\thisrow{linewidth}/1000}, y=mean, col sep=space, y error=stddev]
    {data/nn_16w_17dB.txt};
    \addplot[
    bps_full_8,error bars/.cd,x dir=none, y dir=both, y explicit, 
    ]
    table[x expr={\thisrow{linewidth}/1000}, y=mean, col sep=space, y error=stddev]
    {data/nn_8w_17dB.txt};

     \legend{separated, full-size\cite{rodeGeometricConstellationShaping2022a}, full $n=16$, full $n=8$, square QAM};
   \node [font=\small, anchor=south west] at (axis cs:105,4.5) {$\text{SNR} = \SI{17}{dB}$};
    \end{axis}
    \node[below = 3mm,xshift=-0.7cm] {b)};
    \end{tikzpicture}
\end{subfigure}%
\caption{Performance metrics for $M=64$ with the separated, full-sized, and small demappers}\vspace*{-5ex}%
\label{fig:performance}
\end{figure}
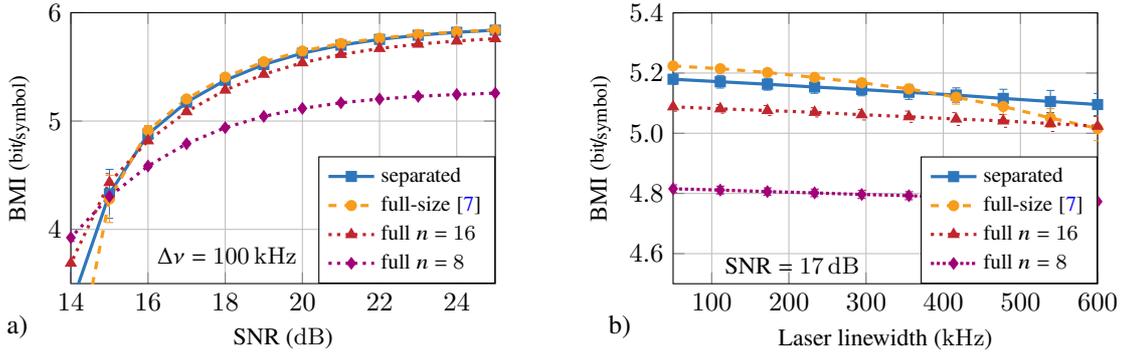

\bibliographystyle{opticajnl}
\bibliography{references.bib}

\end{document}